# Stock Price Prediction Under Anomalous Circumstances

Jinlong Ruan*
*Department of Computer Science*
*University of Rochester*
Rochester, NY, USA
jruan4@u.rochester.edu

Wei Wu*
*Goergen Institute for Data Science*
*University of Rochester*
Rochester, NY, USA
wwu22@u.rochester.edu

Jiebo Luo
*Department of Computer Science*
*University of Rochester*
Rochester, NY, USA
jluo@cs.rochester.edu

*Abstract*—The stock market is volatile and complicated, especially in 2020. Because of a series of global and regional "black swans", such as the COVID-19 pandemic, the U.S. stock market triggered the circuit breaker three times within one week of March 9 to 16, which is unprecedented throughout the history. Affected by the whole circumstance, the stock prices of individual corporations also plummeted by rates that were never predicted by any pre-developed forecasting models. It reveals that there was a lack of satisfactory models that could predict the changes of stocks prices when catastrophic, highly unlikely events occur. To fill the void of such models and to help prevent investors from heavy losses during uncertain times, this paper aims to capture the movement pattern of stock prices under anomalous circumstances. First, we detect outliers in sequential stock prices by fitting a standard ARIMA model and identifying the points where predictions deviate significantly from actual values. With the selected data points, we train ARIMA and LSTM models at the single-stock level, industry level, and the general market level, respectively. Since the public moods affect the stock market tremendously, a sentiment analysis is also incorporated into the models in the form of sentiment scores, which are converted from comments about specific stocks on Reddit. Based on 100 companies' stock prices in the period of 2016 to 2020, the models achieve an average prediction accuracy of 98% which can be used to optimize existing prediction methodologies.

## I. INTRODUCTION

What stock prices will be the next day is the constant question that everybody in the stock market is concerned with. For decades, financial professionals have been trying to explain and forecast the market based on traditional financial models and techniques. Recent years, there have been quite a few stock analysis and price prediction models developed and employed in real-world trading. The majority of these models are based upon an underlying assumption: the stock market has a general behavior pattern. This is usually a plausible assumption because generally people trade stocks with certain patterns and stock prices are backed up by companies' operations and profitability. These models are developed to analyze the data and unveil the underlying pattern, then being utilized to predict stock prices in the future.

However, this assumption does not fully comply with what is observed in the real-world financial markets. One essential factor of the stock market, which is often overlooked by stock price prediction models, is the informational effect of breaking news. For example, on April 1st, 2020, Luckin Coffee Inc.'s stock price went down by 75% due to the disclosure of its financial fraud. Can this event be forecasted by a general stock price prediction model? Likely not. The financial market is full of such kinds of accidents which have huge impacts on stock prices going either way.

Events with a tremendous informational effect on stock prices are deeply connected to real-world business and politics. It is extremely hard to model and predict the occurrences of these events due to the complex and unpredictable nature of life. Nevertheless, these uncertain periods are detectable by tracking the changes of stock prices—when they occur, a general forecasting model's predicted prices usually deviate from the stocks' actual prices by a large amount. Based on the phenomena, this paper first identifies anomalous time periods in a stock's history, discovers a pattern whereby stock prices react to such events during those anomalous periods, and lastly improves the prediction accuracy on the circumstances where general models have unsatisfactory performances.

We provide an unsupervised learning model to improve the stock price prediction accuracy on contextual anomaly time periods. To identify the aforementioned anomalous circumstance, we employ a statistical model, the autoregressive integrated moving average (ARIMA) model, to perform a general prediction on the adjusted closing prices of selected stocks. Next, we analyze the predicted values, locate time periods where ARIMA makes inaccurate predictions and treat them as outliers [1]. With the identified outliers as input data, variants of the statistical models above, seasonal autoregressive integrated moving average models with exogenous variable (SARIMAX), and deep-learning based recurrent neural network models, long short-term memory (LSTM), are trained for the anomalous periods. In support of the models, sentiment data of the matched periods are also calculated and included by the use of VADER sentiment analysis algorithm upon comments from Reddit [2].

The satisfactory results of applying our model on stock price prediction during anomalous time periods show that there exist certain trading patterns in the stock market under the defined anomalous circumstances. Through the comparisons between the prediction results of single-stock prediction models, industrial wide prediction models, and universal models, we suggest that there may exist a generic pattern of how people trade under the defined anomalous circumstances. On a more practical side, our models can be used to enhance the prediction accuracy of general stock price prediction models by taking over the prediction when a stock price prediction model's prediction result is off by a large margin that falls into the defined anomalous circumstances.

The main contributions of this work include:

- We compare two different approaches (parametric and non-parametric) to modeling and predicting stock prices under anomalous circumstances. To the best of our knowledge, there is no existing published study conducted with similar conditions.
- Our models significantly outperform other currently published stock price prediction models in terms of complexity and accuracy.
- Our methodology can be employed to enhance currently available stock predictors' accuracy.
- The experiment results provide circumstantial evidence that stock prices under anomalous circumstances can be modeled sufficiently well by a universal model with the same accuracy as industrial-wide models and single-stock models.

## II. RELATED WORKS

Despite its high volatility and complexity [3], the stock market is one of the most tempting research fields for financial experts and scholars. From statistical methods to artificial intelligence techniques, the toolkit for predicting stock prices and returns has become more and more diverse and well-developed in the past years. Since one primary criterion of a successful stock market prediction is achieving the best result with the least complex model [4], the class of ARIMA models has been a primary choice for most related studies. Through comparison experiments on fifty-six Indian stocks from seven different sectors, Mondal et al. [5] proved that the accuracy of ARIMA models in predicting stock prices is above 85%. Ariyo and Ayo [6] demonstrated that ARIMA is robust in short-term forecasting by using stock data from the New York Stock Exchange and the Nigeria Stock Exchange. Rather than focusing on ARIMA models only, their work also compared the performance of ARIMA to that of other approaches like the artificial neural network (ANN) and the back-propagation neural network (BPNN) in forecasting different stock markets. These works revealed that ARIMA models are comparable to machine-learning-based techniques in time series forecasting.

As one kind of recurrent networks that can choose to forget irrelevant historical signals as predicting the next output, the long-short term memory network (LSTM) has been widely used in natural language processing since it was first proposed by Hochreiter and Schmidhuber [7]. Although researchers seem to prefer other methodologies to LSTM when it comes to the prediction of stock prices, which can be concluded from the number of papers centering around the topic, there are still some notable studies that delve into the applications of LSTM in stock prices forecasting. Using over 170 technical indicators generated by TA-Lib as the input data, Nelson et al. [8], demonstrated that LSTM could reach an average of 55.9% of accuracy at predicting the direction of a particular stock in the next 15 minutes. The work of Chen et al. [9] revealed that LSTM could forecast the earning rate interval of a particular stock in China stock market at an accuracy of 27.2% when the SSE Index (Close, High, Low, Open, Volume) was added into the training and validating process. To measure different models' performances in predicting stock prices, Selvin et al. [10] made a comprehensive comparison between linear (ARIMA) and non-linear algorithms (RNN, LSTM, and CNN). In the task of predicting three companies' stock prices in the next 2000 trading minutes, LSTM achieved an average of 5.3% of error rate, which is much higher than that of the ARIMA model. The experimental results done by Siami-Namini and Namin [11] confirmed the argument that LSTM outperformed ARIMA in predicting adjusted closing prices of particular stocks with an 85% reduction in residual mean square error (RMSE).

Although the studies listed above compare the performance of ARIMA to that of LSTM, none of them integrates sentiment analysis into either of the models as exogenous variables. On the other hand, researchers from a variety of fields have been trying to examine the correlation between public sentiment and the stock market through different approaches. To test the influence of public mood on the closing values of the Dow Jones Industrial Index (DJIA), Bollen et al. [12] compared the prediction of a model solely using past DJIA closing values to those of models combining different mood dimensions extracted from Twitter feeds. The results prove that the inclusion of a specific public mood factor can significantly improve the accuracy of DJIA predictions, which lays the foundation for later studies to explore and develop stock predictive models that take into account public moods. Soon after, Mittal and Goel [13] repeated the work of Bollen et al. but used the k-fold sequential cross validation, which better fits with the changes of financial data, as the performance measure. Their work not only confirms that public moods are indeed predictive of DJIA but also shows that a decent profit could be made by a portfolio management strategy backed up by the model. There are two things worth noting about the aforementioned studies: first, two groups of researchers both built their models based on a Granger Causality Analysis and a Self-Organizing Fuzzy Neural Network; second, the sentiments they extracted and applied are general public moods, which contain both investors' and non-investors' sentiments. Nevertheless, the association between the stock market and sentiments becomes different when other methods and particular sentiments are used. Guo et al. [14] limited the sentiments to the investor sentiment data extracted from a stock-focused social networking platform and applied the Therma Optimal Path (TOP) method to examining the impacts of the sentiments on the index movement in the China stock market. Despite the potential biases caused by the propensities of different stock markets, the results show that the sentiment data can predict the stock price with a relatively high accuracy only when the stock has high investor attention. Interestingly, Jin et al. [23] used social media data about certain popular products to predict the quarterly sales that influence the immediate stock prices around the corporate earning announcements. Their results show that traditional models like the Autoregressive (AR) model and the Bass model can make better predictions on product sales when features extracted from social multimedia platforms are incorporated into the models.

After reviewing aforementioned works that are pertinent to the applications of ARIMA, LSTM, and sentiment analysis in the field of stock price predictions, we design our study in such a way that would distinguish itself from previous studies in the following aspects:

- Extending the "ground truth" of public mood states to a wider online community by adding users from other platforms, like Reddit.
- Incorporating sentiment analysis to both ARIMA and LSTM models.
- Comparing the performances of ARIMA and LSTM models in stock price prediction tasks given the occurrence of abnormal situations and corresponding sentiment scores.

III. METHODOLOGY

This section elaborates on the methods and models that are used to discover selected stock prices' patterns under anomalous circumstances. The main procedure of this research consists of two sub-sections: outlier detection and predictive models training based on detected anomalies.

*A. Stock Data Collection*

To collect stock data, we employ pandas_datareader.data and download 100 stocks in 20 industries from Yahoo Finance by selecting the top five stocks from each sector. The reasons behind this choice consist of two parts: firstly, it is proved that stocks that receive the most attention from investors are more likely to be impacted by investors' sentiments [14]; secondly, top companies undergo a more rigorous audit process which renders their stock prices better represent the companies' real values. Besides, they generally have higher prices and greater trading volumes (i.e., better liquidity). The data of each stock ranges from year 2016 to year 2019, including daily high prices, low prices, closing prices, adjusted closing prices, and volumes. For the interests of this work, we only keep the 3- year daily adjusted closing prices of each stock. The stocks chosen are listed by industries in Table I.

*B. Outlier Detection*

Prior to finding the anomalous time periods, we need to specify what are the anomalous patterns of each stock. Since a general stock price forecasting model is unlikely to detect the occurrence of events with substantial informational effect (which can be considered as randomness), the model is prone to generate predictions that deviate greatly from the actuals under anomalous circumstances. Therefore, we employ ARIMA as our stock price prediction model and identify outliers based on its predicted values' residuals.

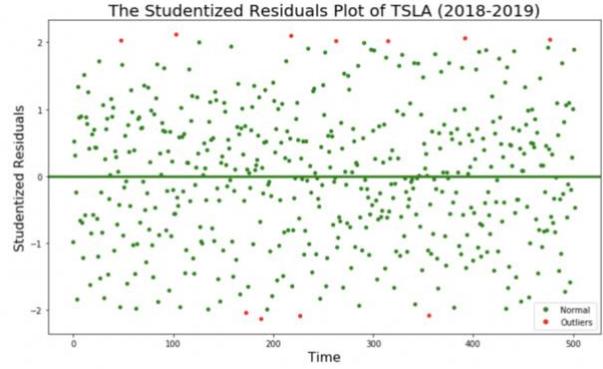

Fig. 1. Single Stock's Flagged Outliers Example

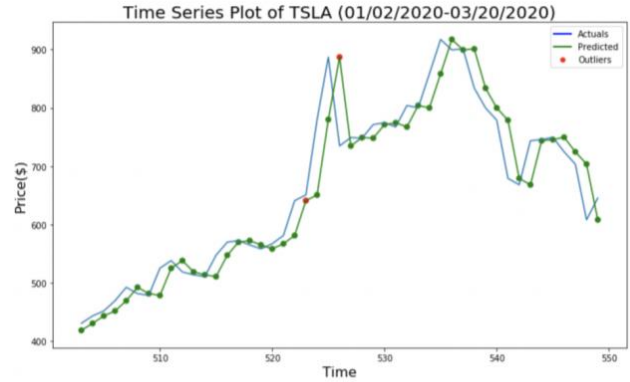

Fig. 2. Outliers in Comparison to Real Trend

We define an anomalous trading day as a trading day when its prediction and actual prices are off by 2 studentized residuals. Therefore, we set the threshold of whether a trading day is flagged as an outlier or not to be 2 studentized residuals. Moreover, when we implement this ARIMA, we set a limit on the sum of hyperparameters to be 3 to reduce the time of fitting the data while ensuring basic performance of the model [5]. We mainly use the adjusted closing prices from year 2017 to train our ARIMA model and let it make predictions in years 2018 and 2019.

However, some stocks' price data in year 2017 cannot be processed by ARIMA due to zero division error. Thus, for these stocks (marked in TABLE I by "*"), we train our ARIMA model based on adjusted closing prices from year 2016 and let it make predictions in years 2018 and 2019. We then analyze the predictions made by ARIMA and flag outliers according to our threshold. As shown in Fig. 1, where the horizontal axis shows the number of trading days that have elapsed since the first trading day in year 2018, the red dots are flagged outliers that deviate greatly from the ground truth. A time series plot that displays the outliers of the same stock caused by more recent catastrophic events is depicted in Fig. 2. The actual dates in Fig. 2 corresponds to the number of trading days that have elapsed since the first trading day in year 2018. We collect all outliers in years 2018 and 2019 as well as their neighboring observations which are located three days before and after the outlier day (7 days in total) to represent the anomalous time periods. In case of errors at training and testing phases, we discard all outliers whose 7-day spans cross two years from the collection of outliers.

TABLE I. 100 Stocks From 20 Industries

| Industries | Airlines | Apparel Retail | Auto Manufactures | Banks | Consumer Electronics | Computer Hardware | Department Stores | Drug Manufacture | Education & Training Services | Entertainment |
|---|---|---|---|---|---|---|---|---|---|---|
| Stock Symbols | UAL | LULU | TSLA | C | AAPL | CAJ | M | GLID* | HLG | NFLX |
| | AAL | CRI* | TM | HSBC | SNE* | ANET | KSS | AMGN | EDU | CHTR |
| | DAL* | TJX | GM | JPM | IRBT | HPQ* | JWN | LLY | STRA | MSG |
| | LUV | GPS | RACE* | BAC | LPL | DELL | DDS | AGN* | GHC* | DIS |
| | ALK | BURL | HMC* | TD | HEAR | WDC | CBD | BIIB | CHGG | DISCA |
| Industries | Financial Data & Stock Exchanges | Gold | Industrial Distribution | Insurance | Medical Instrument & Supplies | Oil & Gas E&P | Railroads | Restaurants | Software-Application | Software-Infrastructure |
| Stock Symbols | NDAQ* | FNV | WSO | HIG* | ISRG* | CNX | CP* | CMG | SHOP | ADBE |
| | FDS | RGLD | FAST | ATH | BDX | COG | NSC | DPZ | FICO | PANW |
| | MSCI | NEM | GWW | AIG | COO | EQT | UNP | MCD | RNG* | MSFT |
| | SPGI | AEM | HDS | ANAT | TFX | XEC | KSU | SBUX | WDAY | VMW |
| | ICE | GOLD | MSM | ESGR | BAX | TPL | CNI | WING | TTD | ORCL |

*C. Sentiment Analysis*

Predictions based solely on historical prices are sometimes erroneous. Since it has been proven by previous works that the incorporation of public moods can significantly improve the predictions about the stock market [12], we introduce sentiment analysis to our models. To collect the sentiment data, we utilize Pushshift Reddit API v4.0 to scrape the comments regarding each stock from Reddit respectively. According to a report from the Pew Research Center, Reddit, as the fifth largest popular site in the U.S., has a userbase accounting for 4% of U.S. adults by 2016 [15]. Of the U.S. adults who use Reddit, 70% of them get news there, which renders a discussion that Reddit is a social news site [16]. As the main information source for the public, news has undoubtedly eminent influences on the stock market. Therefore, we believe that the Reddit users' reactions to stock-related news can represent the public opinions on certain stocks to a reliable extent.

To extract sentiment data from Reddit, we first download all comments in subreddits "stocks", "stock picks", "wallstreetbets", "stockmarket", and "economics" by stock symbols and anomalous periods' dates. Then, we concatenate all the comments from one day and run VADER sentiment analysis on them [2]. For each stock on a specific day, we take the resulting compound score generated by VADER as the sentiment score for the stock on that day. The range of all the sentiment scores is [-1,1], and they are used in the same form for both ARIMA and LSTM models.

It is worth noting that there are a relatively large number of sentiment scores that are 0 since we cannot find any comments on the date pertaining to the stock symbol we specify. There are also some sentiment scores that have relatively large noises since the number of comments on the date and the specified stock symbol is limited. The sentiment scores obtained are not perfect but enough to indicate the public mood towards particular stocks given the purpose of this study.

*D. Find Pattern under Anomalous Circumstances*

At this point, the data we have during anomalous time periods consist of 7-day prices and sentiment scores. Suppose an outlier is at t0, we keep the stock prices as well as the stock's daily sentiment scores from t-3 to t+3 and discard the rest. Given the stock prices and the sentiment scores on t-3, t-2, t-1, and t0, our goal is to predict the stock prices on t+1, t+2, and t+3. In other words, the data of the first four days of each anomaly period are put into the training dataset, while for the rest three days, only the price data are kept for accuracy computation. We apply both ARIMA and LSTM to accomplish this task independently, using anomalous time periods from year 2018 to train the models and testing them on year 2019's anomalous periods. To normalize the data, we calculate the percentage changes in stocks prices each day and use the percentages as the input data instead of the actual stock prices. We compute our model accuracy by converting the predicted percentages changed back to prices and then finding out the L-1 norm accuracy.

*E. ARIMA*

The class of autoregressive integrated moving average (ARIMA) models has been extensively applied to various problems in relation to time series data since they were first proposed by Box et al. [17]. Unlike common regression models, ARIMA models don't suggest assumptions such as strict homoscedasticity and independence between observations [18]. Instead, they incorporate the correlations between current values and their lagged ones as well as forecast errors into the process of predicting future trends while simultaneously capturing the linear relationships between data. Since

ARIMA models are one of statistical approaches, they are also preferable to deep learning methods in terms of interpretability. In this paper, ARIMA models and their extensions are used in two ways: the outlier detection and the prediction of stock prices at pre-identified uncertain periods. As mentioned before, the outlier detection part mainly involves univariate ARIMA models, whose general notation is ARIMA (p,d,q). The p denotes the number of lag observations; the q denotes the size of the moving average window; and the d is the degree of differencing that induces stationarity [19]. The seasonality of stock markets is not taken into consideration at this stage because there is no periodic change observed in either the time series plots of stock prices (using the plot of AAPL as an example) or their correlograms. On the contrary, after the outliers and their associated nodes are detected, a variation of ARIMA models, the seasonal autoregressive integrated moving average models with exogenous factor (SARIMAX), are implemented with a fixed seasonal component s=7 and the sentiment score of a stock at time t. The mathematical expression of the SARIMAX models is [20]:

$$\varphi_p(B)\Phi_P(B^s)\Delta^d\Delta_s^D y_t = \beta x_t + \theta_q(B)\Theta_Q(B^s)\varepsilon_t \quad (1)$$

where:
$y_t$ is the forecast variable
$\varphi_p(B)$ is the non-seasonal AR polynomial of order p
$\theta_q(B)$ is the non-seasonal MA polynomial of order q
$\Phi_p(B^s)$ is the seasonal AR polynomial of order P
$\Theta_Q(B^s)$ is the seasonal MA polynomial of order Q
$\Delta^d$ is the non-seasonal differentiating operator
$\Delta_s^D$ is the seasonal differentiating operator
$\beta$ is the coefficient of the exogenous variable
$x_t$ is the value of the exogenous variable at time t
$\varepsilon_t$ is the white noise

The parameters above are automatically found and tuned by auto.arima function in Pyramid library. We use the percentages changed and the sentiment scores of the first four days in the 7-day span to fit the model, and then the fitted model will make predictions on the percentages changed for the last three days.

*F. LSTM*

Our LSTM model is implemented based on PyTorch neural network library. To inject sentiment analysis into our prediction model, we implement the model as proposed in "Image Captioning at Will: A Versatile Scheme for Effectively Injecting Sentiments into Image Descriptions" by Quanzeng You, Hailin Jin, and Jiebo Luo in 2018. The model implementation detail can be found in their original paper [21]. One thing to notice is that we do not use the loss function proposed with the model since there is no need to compute the loss of sentiment data for the purpose of this paper. Instead, we use the L-1 loss function (i.e., MAE loss function) in out model to compute the loss. Additionally, in our model, we initialize the first sentiment cell's value as 0. We use Adam as our optimizer with default parameters that lr = 0.001, betas = (0.9, 0.999), eps = 1e-08, weight decay = 0, amsgrad = False [22]. The input size and hidden size of our model are both 8.

At the training phase, we train our model by giving it the percentages changed data and sentiment scores on the first four days of the 7-day span and have it output the predictions on the percentages changed of the last three days. Then, we perform backpropagation and gradient descent to update weights of the model against actual percentages changed data of the last three days in the anomalous time period. During the testing phase, we feed the percentages changed data and sentiment scores on the first four days of the span to the model and have it predict the percentages changed for the last three days. Then, we calculate the L-1 norm error and accuracy based on the predictions and actual values. We compute the accuracy of the entire model as the average accuracy over each anomalous period's accuracy.

| Symbols | Date | Outliers | Actuals | Percentage_ | S_Scores |
|---|---|---|---|---|---|
| TSLA | 3/8/18 | 0 | 329.100006 | -0.0096298 | 0.9744 |
| TSLA | 3/9/18 | 0 | 327.170013 | -0.0058645 | 0.7418 |
| TSLA | 3/12/18 | 0 | 345.51001 | 0.05605647 | 0.9582 |
| TSLA | 3/13/18 | 1 | 341.839996 | -0.010622 | 0.889 |
| TSLA | 3/14/18 | 0 | 326.630005 | -0.0444945 | 0.9712 |
| TSLA | 3/15/18 | 0 | 325.600006 | -0.0031534 | 0.9842 |
| TSLA | 3/16/18 | 0 | 321.350006 | -0.0130528 | 0.9444 |
| TSLA | 5/25/18 | 0 | 278.850006 | 0.00359906 | -0.7773 |
| TSLA | 5/29/18 | 0 | 283.76001 | 0.01760805 | 0.7375 |
| TSLA | 5/30/18 | 0 | 291.720001 | 0.02805184 | 0.757 |
| TSLA | 5/31/18 | 1 | 284.730011 | -0.0239613 | 0.9245 |
| TSLA | 6/1/18 | 0 | 291.820007 | 0.02490077 | 0.9752 |
| TSLA | 6/4/18 | 0 | 296.73999 | 0.01685965 | 0.985 |
| TSLA | 6/5/18 | 0 | 291.130005 | -0.0189054 | 0.9985 |

Fig. 3. Head Rows of Training Dataset

| Symbols | Date | Outliers | Actuals | Percentage_ | S_Scores | Predicted Price |
|---|---|---|---|---|---|---|
| TSLA | 1/15/19 | 0 | 344.429993 | 0.02999402 | 0.7894 | N/A |
| TSLA | 1/16/19 | 0 | 346.049988 | 0.00470341 | 0.9984 | N/A |
| TSLA | 1/17/19 | 0 | 347.309998 | 0.00364112 | 0.8295 | N/A |
| TSLA | 1/18/19 | 1 | 302.26001 | -0.1297112 | -0.883 | N/A |
| TSLA | 1/22/19 | 0 | 298.920013 | -0.0110501 | 0.9833 | 308.9663321 |
| TSLA | 1/23/19 | 0 | 287.589996 | -0.0379032 | 0.9962 | 300.4864369 |
| TSLA | 1/24/19 | 0 | 291.51001 | 0.01363056 | 0 | 285.2844527 |
| TSLA | 4/1/19 | 0 | 289.179993 | 0.03330239 | 0.7768 | N/A |
| TSLA | 4/2/19 | 0 | 285.880005 | -0.0114115 | 0.9602 | N/A |
| TSLA | 4/3/19 | 0 | 291.809998 | 0.02074294 | 0 | N/A |
| TSLA | 4/4/19 | 1 | 267.779999 | -0.0823481 | -0.9062 | N/A |
| TSLA | 4/5/19 | 0 | 274.959992 | 0.02681303 | 0 | 270.4434873 |
| TSLA | 4/8/19 | 0 | 273.200012 | -0.0064009 | -0.941 | 274.1737958 |
| TSLA | 4/9/19 | 0 | 272.309998 | -0.0032577 | 0 | 269.0768449 |

Fig. 4. Sample Result of Predictions

IV. EXPERIMENT RESULTS

Fig. 3 includes head rows of the training dataset that is fed into the aforementioned LSTM and SARIMAX models. Fig. 4 includes head rows of testing dataset after prediction. Since the goal of this study is to compare the performances of LSTM and SARIMAX in forecasting stocks' adjusted closing prices under anomalous circumstances, each class of models is trained by training datasets of three different scales, which ends up with three categories of models for each stock. Respectively, they are:

- Universal model: the forecasting model trained with past outliers and neighboring observations of all stocks
- Industry-wide model: the forecasting model trained with past outliers and neighboring observations of stocks from the same industry
- Single-stock model: the forecasting model trained with the target stock's own past outliers and neighboring observations

For example, to obtain the SARIMAX models for TSLA, we use a dataset of all stocks, a dataset of the stocks from the automobile industry, and a dataset of TSLA to train a model, respectively, so that we have three SARIMAX models for TSLA in the end. Likewise, we obtain three LSTM models for TSLA, so there are six forecasting models for TSLA. The SARIMAX and LSTM models trained by the dataset of all stocks are categorized as the universal model, the models trained by the dataset of the stocks from the automobile industry are categorized as the industry-wide model, and the models trained by the dataset of TSLA are categorized as the

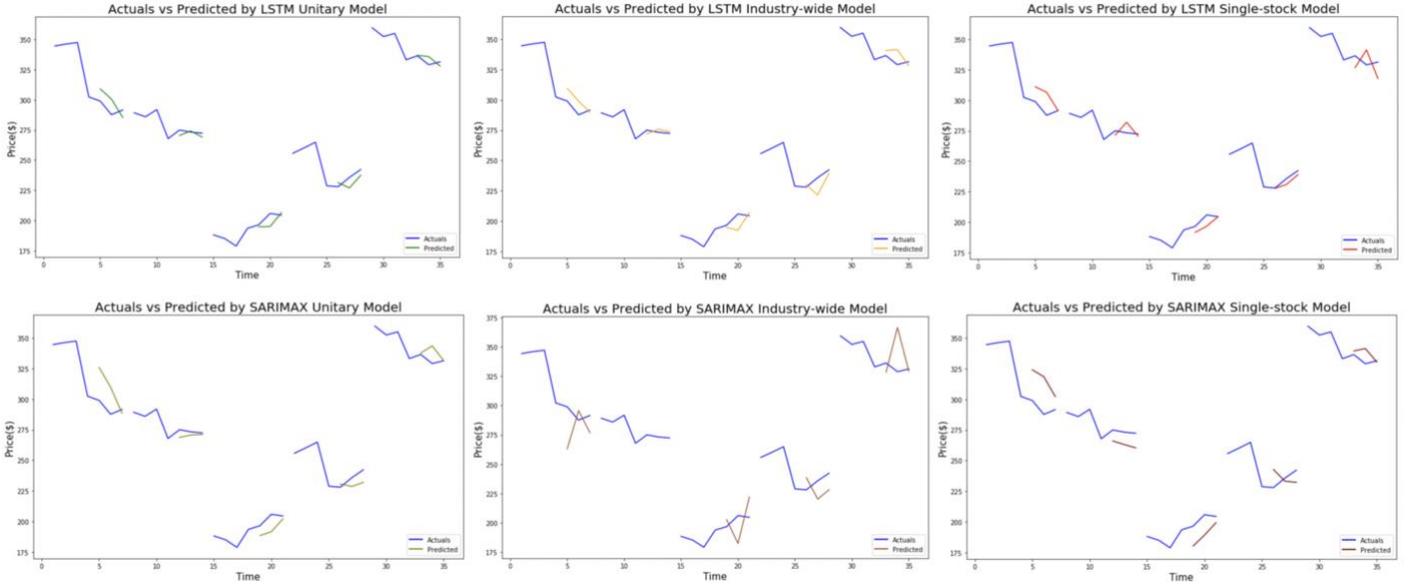

Fig. 5. TSLA: Five Anomalous Time Periods in 2019

TABLE II. LSTM Models Trained on Epochs 10-1000

| Models | LSTM (Epoch=10) | | | LSTM (Epoch=100) | | | LSTM (Epoch=1000) | | |
|---|---|---|---|---|---|---|---|---|---|
| | Universal Model | Industry-wide Model | Single-stock Model | Universal Model | Industry-wide Model | Single-stock Model | Universal Model | Industry-wide Model | Single-stock Model |
| Average Accuracy (%) | 98.59 | 98.38 | 90.08 | 98.69 | 98.44 | 98.21 | 98.61 | 98.25 | 97.99 |
| Time (sec) | 13 | 13 | 12 | 57 | 59 | 64 | 696 | 726 | 734 |

TABLE III. LSTM vs. SARIMAX

| Models | LSTM (Epoch=100) | | | SARIMAX | | |
|---|---|---|---|---|---|---|
| | Universal Model | Industry-wide Model | Single-stock Model | Universal Model | Industry-wide Model | Single-stock Model |
| Average Accuracy (%) | 98.69 | 98.44 | 98.21 | 97.77 | 98.15 | 98.14 |
| Time (sec) | 57 | 59 | 64 | 46567 | 1219 | 974 |

single-stock model. Next, as shown in Table II, the LSTM models with different epochs are tested in order to find the one that best exemplifies the performances of LSTM in each scenario. We calculate our result accuracy by MAPE. The specific formula for a single trading day prediction accuracy is given below:

$$\text{Accuracy} = 1 - \frac{1}{n}\sum_{i=1}^{n}\left|\frac{A_i - F_i}{A_i}\right| \times 100\% \quad (2)$$

where:
$A_i$ is the actual stock price
$F_i$ is the predicted stock price
$n$ is the number of predictions

The overall average accuracy and the time are calculated by averaging the accuracies and time of the models of all stocks. Except for the single-stock model at epoch=10, the models with epoch=100 and epoch=1000 show no significant difference in accuracy. In addition, Siami-Namini et al. [11] demonstrated in their work that an increasing number of training rounds (epochs) on the same data does not improve the accuracy of an LSTM forecasting model, so in this paper, only the class of LSTM models with an epoch of 100 is used to compare with SARIMAX models from the same category.

According to Table III, the difference between average accuracies of LSTM and SARIMAX models is minor from either a holistic or a regional perspective. Nevertheless, as the amount of training data increases (i.e. the size of the training dataset of the single-stock model is the smallest, that of the industry-wide model the second, the that of the universal model is the smallest), the class of SARIMAX models falls into a competitive disadvantage: its running time grows at a geometric rate. In comparison, the running time of LSTM models is relatively constant, which reinforces the argument that LSTM outperforms SARIMAX in handling enormous data.

In terms of the specific performances of both models, a series of plots in Fig. 5 speak for themselves. Take TSLA as an instance: its stock prices changed anomalously five times in 2019, which are represented by five separate lines in the plots. The blue lines denote the actual adjusted closing prices, and the lines in other colors denote the predicted adjusted closing prices generated by different models. As demonstrated, the prices predicted by LSTM better coincide with the real prices and trends than those of SARIMAX do; moreover, the SARIMAX industry-wide model even fails to forecast the adjusted closing prices for the second period, which further diminishes its effectiveness on this task.

## V. CONCLUSION

In this paper, we employ two different models, LSTM and ARIMA, to predict stock prices under anomalous circumstances. Both of these models result in satisfactory prediction accuracy. However, as demonstrated in the experiment results, LSTM outperforms ARIMA in terms of time complexity and the ability to follow the trends of stock prices. There are several important implications by our findings. First of all, our results generate the evidence of the existence of a movement pattern of stock prices under anomalous circumstances, and it is likely to model the pattern and perform predictions with an accuracy as high as 98%. Moreover, this pattern is potentially applicable to all stocks in the U.S. stock market and can be captured by a universal model. This may imply that different people who trade in different sectors of the U.S. stock market have a generic or universal trading pattern that can be modeled by big data approaches. A more practical application of our models can be improving the prediction results of general stock price prediction models when a defined anomalous period happens. When a bad prediction that falls into particular anomalous periods occurs in general stock price prediction models, our model can predict future stock prices based on the information of the outlier day and three trading days before the outlier day. This can greatly improve the prediction accuracy over the next three trading days.

At present, since our research is limited to 100 stocks due to the constraint of computing resources, it is not guaranteed that the observed pattern will still hold true if a broader range of stocks is taken into account. A future research direction will focus on incorporating more data from other stocks and stock markets, or more data from other sources (including social media signals [23] [24]), into model training and optimization. Furthermore, the sentiment data in this study is extracted from the Reddit comments only, so the power of a more general sentiment analysis in improving predictive models of the stock market remains unclear and needs further research. We will also integrate our models with other existing stock prediction methodologies and improve the models' overall performance. At present, we cannot interpret our model – what is inside the black box still remains unknown—therefore we may also work on interpreting our models in the future work.

## VI. ACKNOWLEDGMENT

This research is supported in part by the New York State CoE Goergen Institute for Data Science at the University of Rochester and a generous gift from the Jefferies Financial Group.